\documentstyle[aps,epsf,twocolumn]{revtex}

\begin{document}
\title{$c$-Axis Superfluid Response and Pseudogap in High-$T_c$ Superconductors}

\author{C. Panagopoulos$^1$, J.R. Cooper$^1$, T. Xiang$^2$, Y.S. Wang$^3$ and C.W. Chu$^3$}

\address{$^1$ Cavendish Laboratory and Interdisciplinary Research Centre in Superconductivity, University of Cambridge, Madingley Road, Cambridge CB3 0HE, UK}

\address{$^2$ Institute of Theoretical Physics, Academia Sinica, P.O. Box 2735, Beijing 100080, PRC}

\address{$^3$ Department of Physics and Texas Centre for Superconductivity at the University of Houston, Houston, Texas 77204-5932, USA}

\maketitle

\begin{abstract}
To gain insight into the out-of-plane electrodynamics of high-$T_c$ superconductors we have measured the absolute values and temperature dependence of the $c$-axis magnetic penetration depth $\lambda _c(T)$ for two typical single layer high-$T_c$ cuprates, La$_{2-x}$Sr$_x$CuO$_4$ and HgBa$_2$CuO$_{4+x}$ as a function of doping. A distinct change in the behaviour of $\lambda _c$ is observed near 0.20 holes/Cu, which is related to the opening of the normal state pseudogap. The variation of $\lambda _c$ with doping is qualitatively similar to that of the in-plane component, $\lambda _{ab}$, which suggests that the $c$-axis superfluid response is mainly governed by the superconducting condensation energy. The strong doping dependence of $\lambda _c(0)$ for $p<0.20$ provides an explanation for the discrepancies in the literature.

\end{abstract}

\vspace{1cm}
 
The $c$-axis magnetic penetration depth $\lambda _c$ in high-$T_c$ superconductors (HTS) is a key parameter for understanding the interlayer dynamics of electrons. It is also important for assessing the proposed pairing mechanism arising from in-plane confinement, namely the interlayer tunneling (ILT) theory which was proposed by Anderson and co-workers \cite{JMW,PWA1,SC1,PWA2}. Although great effort has been devoted to the accurate measurement of $\lambda _c$, the experimental results are still controversial. For example, for optimally doped La$_{2-x}$Sr$_x$CuO$_4$, optical \cite{Uchida} and surface impedance \cite{Shibauchi} measurements found $\lambda _c(T=0K)\simeq 4\mu m$, in agreement with the prediction of ILT theory. On the other hand for Tl$_2$Ba$_2$CuO$_{6+x}$, vortex imaging \cite{Moler1} and optical \cite{Tsetkov} measurements yielded $\lambda _c(0)\simeq 19\mu m$, and more recent optical measurements gave $12\mu m$ \cite{BasovTl}. These high values of $\lambda _c(0)$ cannot be easily accounted by the ILT model \cite{SC2}. Furthermore, for nearly optimally doped HgBa$_2$CuO$_{4+x}$ ($T_c=96K$), the values of $\lambda _c(0)$ obtained by vortex imaging \cite{Kirtley} and optical \cite{BasovHg} measurements are $\simeq$ $8\mu m$ and $6\mu m$, respectively. However, $ac$-susceptibility measurements for slightly overdoped HgBa$_2$CuO$_{4+x}$ ($ T_c$=93K) give $\lambda _c(0)\simeq 1.5\mu m$ \cite{PanaHg} in agreement with ILT theory. The large discrepancies in the reported values of $\lambda _c(0)$ for HTS with one CuO$_2$ plane per unit cell makes it difficult to reach a consensus regarding the trends of $\lambda _c(0)$ in HTS and also the validity of the ILT model \cite{SC2,AJL,PWA3,Farid}. Moreover, the strong doping dependence of both the superconducting and normal state properties \cite{Timusk} and the presence of the pseudogap in under- and optimally doped samples may play a significant role in the behaviour of the $c$-axis electrodynamics. It is therefore important to study the $c$-axis superfluid response as a function of carrier concentration and search for possible correlations between the in-plane and out-of-plane superconducting properties and the normal state pseudogap.

In this paper we report high quality data for $\lambda _c$, measured by the $ac$-susceptibility technique, as functions of doping and temperature for La$_{2-x}$Sr$_x$CuO$_4$ (LSCO) and HgBa$_2$CuO$_{4+x}$ (Hg-1201). We find that $\lambda _c(0)$ changes rapidly with doping when $p<0.20$, and is almost independent of doping for $p\geq 0.20$, where $p$ is the hole concentration per planar Cu atom. A crossover near $p=0.20$ is also observed in the behaviour of the temperature dependence of $\lambda _c^{-2}$, similar to that previously reported for the in-plane counterpart, $\lambda _{ab}^{-2}$ \cite{PanaLSC}, suggesting that it is also strongly affected by the presence of the normal state pseudogap. The doping dependence of $\lambda _c(0)$ shown here lends some support to the ILT model in the sense that it can account for the discrepancies in the reported values of $\lambda _c(0)$ for Hg-1201.

The samples studied here were magnetically aligned high quality single-phase powders of La$_{2-x}$Sr$_x$CuO$_4$ ($p=x=$ 0.10, 0.15, 0.20, 0.22 and 0.24) \cite{PanaLSC} and HgBa$_2$CuO$_{4+x}$ ($x$=0.10, $i.e.$, $p$=0.09, with $T_c$=60 K) \cite{Xiong}. Measurements were performed down to $1.2K$ for LSCO and $4.2K$ for Hg-1201 using the low-field $ac$-susceptibility technique with an $ac$-field of $1G$ $RMS$ at $333Hz$ parallel to the CuO$_2$ planes. For LSCO the value of $p$ is taken to be equal to the Sr content $x$, whereas for Hg-1201 it was determined by estimating the oxygen content and verified by thermopower measurements as in ref. \cite{Xiong}. Transport, magnetic and spectroscopic measurements confirmed the high quality of our samples. The data were analysed using London's model for small aligned crystallites of anisotropic HTS. Details of the technique and data analysis can be found elsewhere \cite{Porch,PanaHg}. The low-field $ac$-susceptibility technique gives the fractional diamagnetism of a superconductor and thus is a direct probe of the magnetic penetration depth \cite{PanaHg}. For each doping concentration of LSCO, the result presented here is typical of those obtained for four samples prepared independently from the same polycrystalline pellet. Each of these samples was aligned at different times and from each sample 4 pieces were cut and measured. Therefore a total of 16 samples were investigated for each doping content. In the case of Hg-1201, two samples were prepared from the same polycrystalline pellet and again 4 pieces of each aligned sample were cut and measured for each doping level. The error of $\lambda _c(0)$ is estimated to be $<$ $8\%$ for LSCO and $<$ $10\%$ for Hg-1201, while the error in the relative temperature dependence $\lambda _c(T)$/$\lambda _c(0)$ is $<$ $1\%$ and is therefore negligible.

Fig. 1(a) shows the doping dependence of $\lambda _c(0)$ for LSCO and Hg-1201. Lets us first discuss the case of LSCO. For comparison our previously published data of $\lambda _{ab}(0)$ for the same LSCO samples are also included \cite{PanaLSC}. The ratio of $\lambda_c(0)$/$\lambda _{ab}(0)$ measures the electromagnetic anisotropy $\gamma$ of the system. The values of $\gamma $ for the samples we measured are shown in Fig 1(b) and compared with those obtained by Shibauchi $et$ $al.$ for LSCO single crystals with $x<0.20$ using the surface impedance technique \cite{Shibauchi}. There is very good agreement between the two sets of data despite the large error bars in Fig. 2(a) of ref. \cite{Shibauchi}. From Fig. 1 we find that for the intermediate doping regime, $0.14<x<0.20$, $T_c$ does not change very much, although there are significant changes in $\lambda _c(0)$ and $\gamma$. Namely, $\lambda _c(0)$, $\lambda _{ab}(0)$ and $\gamma$ increase rapidly for $x=p<0.20$ holes/Cu and are nearly independent of doping for $x=p\geq 0.20$ with $\lambda _c(0)$ $\simeq 2\mu m$ and $\gamma$ $\simeq 12$.

Figure 2(a) shows the temperature dependence of the $c$-axis superfluid density of LSCO as a function of doping. There is a noticeable change in the curvature of $\lambda _c^{-2}(T)$ near $0.20$ holes/Cu as previously observed in  $\lambda_{ab}^{-2}(T)$ and shown in Fig. 2(b) for comparison \cite{PanaLSC}. The relatively sudden change in curvature as well as in the absolute values of $\lambda _{ab,c}^{-2}$ near $x=0.20$ is not a feature unique to the superfluid response. It has also been observed in several other physical quantities, both in the normal and superconducting states, and attributed to the opening of the pseudogap in the low doping regime \cite{Loram94,Ito,Ding,Norman,Boebinger}. The physical mechanism leading to the normal state pseudogap is still unknown. However, it is generally believed to cause a loss of the normal state spectral weight near the Fermi energy \cite{PanaLSC,Loram94,Loram}. The loss of spectral weight is expected to strongly suppress the superfluid density $\lambda _{ab,c}^{-2}$ and alter the temperature dependence of both $\lambda _{ab}$ and $\lambda _c$ \cite{Loram,Lee,Chen,Williams}.

By comparing the data of $\lambda _c(0)$ for LSCO with those for Hg-1201, we find that the doping dependences of $\lambda _c(0)$ in these two compounds are very similar [Fig. 1(a)]. Namely, for an underdoped, an almost optimally doped (measured independently by two different techniques), and two overdoped Hg-1201 samples, with $p\sim $ {0.09 [$this$ $work$], 0.16 \cite{Kirtley} and \cite{BasovHg}, 0.18 \cite{PanaHg} and 0.23 \cite{BasovCP} $\pm0.01$} and $T_c$ = ($60K$, $96K$ \cite{Kirtley}, $96K$ \cite{BasovHg}, $93K$ \cite{PanaHg} and $60K$ \cite{BasovCP}), the values of the $c$-axis penetration depth are $\simeq (8.5\mu m$, $8 \pm1 \mu m$ \cite{Kirtley}, $6 \pm1 \mu m$ \cite{BasovHg}, $1.5\mu m$ \cite{PanaHg} and $2\mu m \cite{BasovCP})$, respectively. These results are shown in Fig. 1(a) together with the LSCO data for comparison. The doping content for Hg-1201 was estimated using the universal relation $T_c$=$T_{c,max}$[1-82.6($p$-0.16)$^2$], where the maximum is reached for an optimum doping of $p$ $\simeq 0.16$ holes per CuO$_2$ plane \cite{TallonTcp}. In the case of Hg-1201 $T_{c,max}$ is chosen to be $96K$. For the samples studied here and in refs \cite{PanaHg,BasovCP} the values of $p$ estimated were confirmed from thermoelectric power measurements as in ref. \cite{Xiong}. Although there is some uncertainty in determining the value of $p$ in Hg-1201, these results indicate that the doping dependence of $\lambda _c(0)$ is qualitatively similar for these two single-layer high-$T_c$ cuprates.

Our results show that there is no experimental inconsistency in the wide range of $\lambda _c(0)$ values reported in the literature for the Hg-1201 compound. Namely, the value of $1.5\mu m$ that we obtained \cite{PanaHg} for a slightly overdoped sample by the $ac$-susceptibility technique corresponds to the simple situation where there is no normal state gap. Although the values of Kirtley $et$ $al$. \cite{Kirtley} and Basov $et$ $al$.  \cite{BasovHg} correspond to samples with high values of $T_c$ ($96K$) the results shown here strongly suggest that their values of $\lambda _c(0)$ correspond to significantly lower doping levels where the normal state gap comes into play. In this sense our results support the ILT model- since the only clear exception to this model is now the Tl-2201 compound \cite{Moler1,Tsetkov,BasovTl}. However, we should note that for any pairing mechanism (not just ILT) the presence of the normal state gap also leads to a similar correlation between the condensation energy, and the superfluid density, $i.e.$, $\lambda _c(0)$ and $\lambda _{ab}(0)$ \cite{Loram94,Loram}. There are also distinct differences between $\lambda _c$ and $\lambda _{ab}$. The anisotropy $\gamma$ increases on the underdoped side and the $T$-dependences at low temperatures (at $T$ $<$ $0.3T_c$) are very different. Namely, the in-plane component varies always linearly with $T$ whereas the $c$-axis counterpart varies as $T^n$ with $n \simeq 3$ for LSCO [$this$ $work$] and $n \simeq 5$ for Hg-1201 \cite{PanaHg}. Details on possible mechanisms leading to these power laws can be found in ref. \cite{Taorev}. Therefore, it is not immediately obvious that the qualitative similarities between $\lambda _c(T)$ and $\lambda _{ab}(T)$ shown in Fig. 2 can be understood within the ILT theory where interplane Josephson effects dominate the $c$-axis electrodynamics.

In summary, we have measured the effects of carrier concentration on the $c$-axis superfluid response of two typical monolayer high-$T_c$ cuprates. We observe a crossover in the behaviour of $\lambda _c(0,T)$ near $p=0.20$, which is closely related to the opening of the pseudogap. The strong doping dependence of $\lambda _c(0)$ for $p<0.20$ provides an explanation for the discrepancies in the literature. The qualitatively similar behaviour of $\lambda _c$ and $\lambda _{ab}$ as a function of doping suggests that both quantities are strongly affected by the presence of the normal state pseudogap, as is the superconducting condensation energy.

We thank D.N. Basov, J.R. Kirtley, K.A. Moler, J.W. Loram and P.W. Anderson for useful discussions in the past year. C.P. thanks Trinity College, Cambridge for financial support through a research fellowship.

\begin{figure}
\leavevmode\epsfxsize=8.5cm
\epsfbox{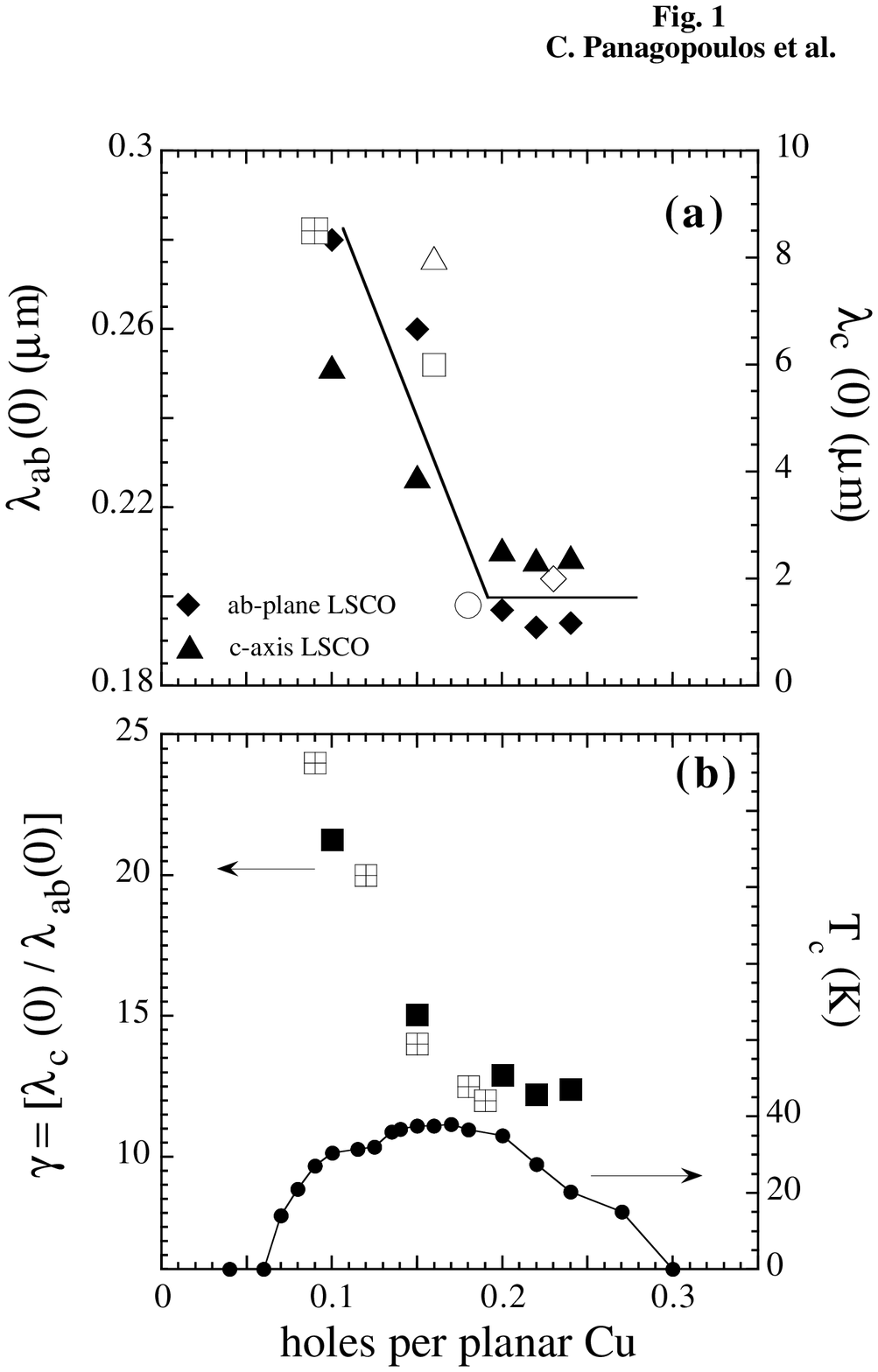}
\caption{
(a) Absolute values of the in-plane, $\lambda_{ab}$, (closed diamonds) and out-of-plane, $\lambda_c$, (closed triangles) magnetic penetration depth as a function of doping for La$_{2-x}$Sr$_x$CuO$_4$. The $\lambda_{ab}(0)$ data are taken from ref. [18]. $\lambda_c(0)$ data for HgBa$_2$CuO$_{4+x}$ are also included for $p\sim $ 0.09 (crossed square) [$this$ $work$], 0.16 (open triangle) [11], 0.16 (open square) [12], 0.18 (open circle) [13] and 0.23 (open diamond) [30]. The doping dependence of $\lambda_{ab}(0)$ in Hg-1201 is similar to that of LSCO [18]. The solid line is drawn as a guide to the eye. (b) Doping dependence of the anisotropic ratio $\lambda_c(0)$/$\lambda_{ab}(0)$ for La$_{2-x}$Sr$_x$CuO$_4$. The closed circles represent the values of $T_c$ and the closed squares the values of $\gamma$ measured here. Crossed squares are values of $\gamma$ taken from Shibauchi $et$ $al.$ [6] for comparison.
}
\end{figure}

\begin{figure}
\leavevmode\epsfxsize=8.5cm
\epsfbox{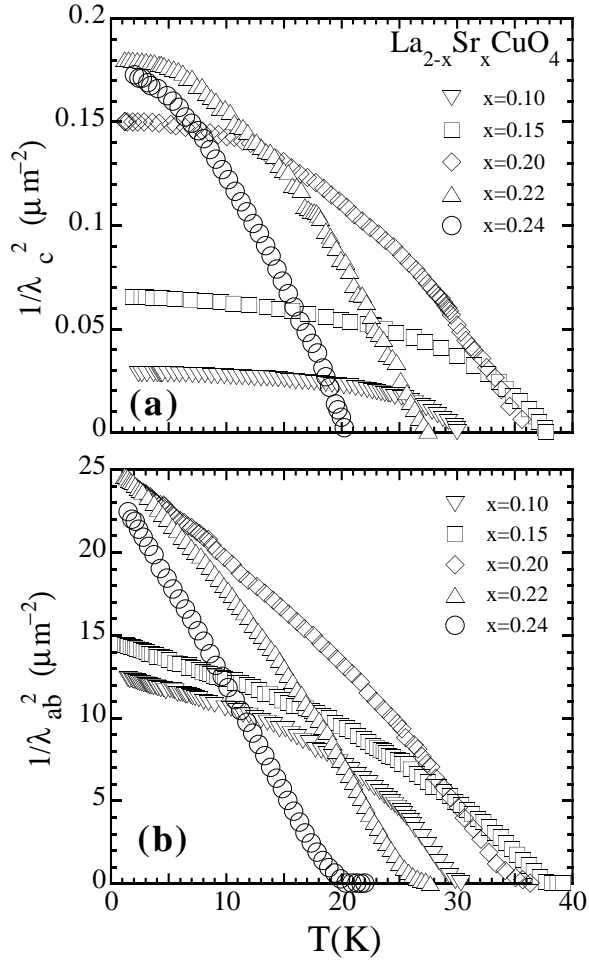}
\caption{
(a) Temperature dependence of the $c$-axis superfluid response, $\lambda_c^{-2}$, as a function of doping for La$_{2-x}$Sr$_x$CuO$_4$. (b) Temperature dependence of the in-plane superfluid response, $\lambda _{ab}^{-2}$, as a function of doping for the samples shown in panel (a). The data in this panel are taken from ref. [18].
}
\end{figure}


\begin{references}
\bibitem{JMW}  J.M. Wheatley, T. Hsu and P.W. Anderson, Nature {\bf 333}, 121 (1988).


\bibitem{PWA1}  P.W. Anderson, Science {\bf 256}, 1526 (1992).


\bibitem{SC1}  S. Chakravarty, A. Sudbo, P.W. Anderson and S. Strong, Science {\bf 261}, 337 (1993).


\bibitem{PWA2}  P.W. Anderson, Science {\bf 268}, 1154 (1995).


\bibitem{Uchida}  S. Uchida, K. Tamakasu and S. Tajima,  Phys. Rev. B {\bf 53}, 14558 (1996).


\bibitem{Shibauchi}  T. Shibauchi $et$ $al.$, Phys. Rev. Lett. {\bf 72}, 2263- (1994).


\bibitem{Moler1}  K.A. Moler, J.R. Kirtley, D.G. Hinks, T.W. Li and M. Xu, Science {\bf 279}, 1193 (1998).


\bibitem{Tsetkov}  A.A. Tsetkov $et$ $al.$, Nature {\bf 395}, 360 (1998).


\bibitem{BasovTl}  D.N. Basov $et$ $al.$, Science {\bf283}, 49 (1999).


\bibitem{SC2}  S. Chakravarty, H-Y. Kee and E. Abrahams,  Phys. Rev. Lett. {\bf 82}, 2366 (1999).


\bibitem{Kirtley}  J.R. Kirtley $et$ $al.$, Phys. Rev. Lett. {\bf 81}, 2140 (1998).


\bibitem{BasovHg}  D.N. Basov $et$ $al.$, ($unpublished$).


\bibitem{PanaHg}  C. Panagopoulos, J.R. Cooper, T. Xiang, G.B. Peacock, I. Gameson and P.P. Edwards, Phys. Rev. Lett. {\bf 79}, 2320 (1997).


\bibitem{AJL}  A.J. Leggett, Science {\bf 279}, 1157 (1998).


\bibitem{PWA3}  P.W. Anderson,  Science {\bf 279}, 1196 (1998).


\bibitem{Farid}  B. Farid, J. Phys.: Condens. Matter {\bf 10}, L589 (1998).


\bibitem{Timusk}  T. Timusk and B. Statt, Rep. Prog. Phys. {\bf 62}, 61 (1999).


\bibitem{PanaLSC}  C. Panagopoulos $et$ $al.$, Phys. Rev. B ($in$ $press$); also at cond-mat/9903117.


\bibitem{Xiong}  Q. Xiong $et$ $al.$, Phys. Rev. B {\bf 50}, 10346 (1994).


\bibitem{Porch}  A. Porch $et$ $al.$,  Physica C, {\bf214}, 3508 (1993).


\bibitem{Loram94}  J.W. Loram $et$ $al.$,  Physica C, {\bf235-240}, 134 (1994).


\bibitem{Ito}  T. Ito,  K. Takenaka and S. Uchida,  Phys. Rev. Lett. {\bf 70}, 3995 (1993).


\bibitem{Ding}  H. Ding $et$ $al.$, Phys. Rev. Lett. {\bf 78}, 2628 (1997).


\bibitem{Norman}  M.R. Norman $et$ $al.$,  Nature {\bf 392}, 157 (1998).


\bibitem{Boebinger}  G. Boebinger $et$ $al.$, Phys. Rev. Lett. {\bf 77}, 5417 (1998).


\bibitem{Loram}  J.W. Loram, K.A. Mirza and J.R. Cooper, "High Temperature Superconductivity", research review (editor: Liang, W.Y., University of Cambridge, 1998).


\bibitem{Lee}  P.A. Lee and X.-G. Wen,  Phys. Rev. Lett. {\bf78}, 4111-4114 (1997).


\bibitem{Chen}  Q. Chen, I. Kosztin, B. Janko and K. Levin, Phys. Rev. Lett. {\bf81}, 4708 (1998).


\bibitem{Williams}  G.V.M. Williams, E.M. Haines and J.L. Tallon, Phys. Rev. B {\bf57}, 146 (1998).


\bibitem{BasovCP}  E.J. Singley and D.N. Basov ($private$ $communication$).


\bibitem{TallonTcp}  J.L. Tallon $et$ $al.$, Phys. Rev. B {\bf51}, 12911 (1995).  


\bibitem{Taorev}  T. Xiang, C. Panagopoulos and J.R. Cooper, Int. J. Mod. Physics B {\bf12}, 1007 (1998).

\end{references}
\end{document}